\documentclass [11pt, letterpaper]{article}
\pdfoutput=1 
\usepackage{fullpage}
\usepackage{amsmath}
\usepackage{amssymb}
\usepackage{graphicx}
\usepackage{braket}
\usepackage{mathrsfs}
\usepackage{wrapfig}
\usepackage{boxedminipage}
\usepackage{epsfig}
\usepackage{setspace}
\usepackage{subfigure}
\usepackage{float}
\usepackage{hyperref}
\usepackage{enumerate}
\usepackage{cite}
\usepackage[font=footnotesize,labelfont=rm]{caption}
\usepackage{color}
\usepackage{xcolor}

\def\dbar{{\mathchar '26\mkern-12mu d}}

\begin{document}

\begin{flushright}
{\tt arXiv:1806.05170}
\end{flushright}


\bigskip
\bigskip

\bigskip
\bigskip
\bigskip
\bigskip

\begin{center} 

{\Large\bf  Holographic Heat Engines,  Entanglement Entropy,}

\bigskip
\bigskip

{\Large\bf  and }

\bigskip
\bigskip

{\Large\bf   Renormalization Group Flow} 

\end{center}

\bigskip \bigskip \bigskip \bigskip

\centerline{\bf Clifford V. Johnson and Felipe Rosso}

\bigskip
\bigskip
\bigskip

\centerline{\it Department of Physics and Astronomy }
\centerline{\it University of
Southern California}
\centerline{\it Los Angeles, CA 90089-0484, U.S.A.}

\bigskip

\centerline{\small {\tt johnson1} and {\tt felipero} [at] usc [dot] edu}

\bigskip
\bigskip


\begin{abstract} 
\noindent We explore a fruitful connection between the physics of conformal field theories (CFTs) in $d$--dimensional Minkowski spacetime and the extended gravitational thermodynamics of hyperbolic black holes in $(d+1)$--dimensional anti--de Sitter spacetime. The CFTs are reduced on a  region bounded by a sphere. We show that Renormalization Group  flows between CFTs are specific thermodynamic processes in the $(p,V)$ plane, where the irreversibility of coarse--graining flows from the ultraviolet  to the infrared  corresponds to the Second Law of thermodynamics, preventing heat from flowing from low temperature to high. We observe that holographic heat engines using the black holes as a working substance correspond to specific combinations of CFT flows and deformations. We construct three special engines whose net heat and  work can be described in terms of changes of entanglement entropy across the sphere. Engine efficiencies emerge as simple functions of the ratio of the number of degrees of freedom of two CFTs. 
\end{abstract}

\pagenumbering{gobble}

\newpage 

\pagenumbering{arabic}

\baselineskip=18pt 
\setcounter{footnote}{0}

\section{Introduction}
\label{sec:introduction}
In the last two decades there has been a lot of progress in the understanding of  Renormalization Group (RG) flows in quantum field theories, using entanglement entropy as a diagnostic tool. The ability to compute the entanglement entropy in higher dimensional quantum field theories  using geometric techniques\cite{Ryu:2006bv,Ryu:2006ef} with holographic dualities\cite{Maldacena:1997re, Gubser:1998bc, Witten:1998qj, Witten:1998zw}  has been crucial. In refs.\cite{Myers:2010xs,Myers:2010tj} it was conjectured that the universal contribution to the entanglement entropy of a sphere is a decreasing quantity along the RG flow. For a general conformal field theory (CFT) in $d$ spacetime dimensions possessing a $(d+1)$--dimensional gravitational dual that is asymptotically anti--de Sitter (AdS), this universal contribution was shown to be proportional to (for Einstein gravity) \cite{Casini:2011kv}\footnote{See {\it e.g.} ref.~\cite{Myers:2010tj} for analogous expressions in other kinds of gravity theory.}:
\begin{equation}\label{eq:a-function}
a^*_d=\frac{\pi^{d/2-1}}{8 \Gamma(d/2)}\frac{L^{d-1}}{G}\ ,
\end{equation}
where $L$ is the AdS radius and $G$ is Newton's constant. Their precise ratio in eq.~(\ref{eq:a-function}) can be written in terms of purely field theory quantities using the holographic duality. For even values of $d$ this quantity matches the coefficient of the A--type trace anomaly, meaning that in this case the conjecture coincides with an earlier conjecture by Cardy\cite{Cardy:1988cwa}. The claim was also in accord with the entanglement entropy proof of the $c$--theorem\cite{Casini:2004bw}. Casini~{\it et.~al.} were able to additionally show that this conjecture is indeed correct for three and four \cite{Casini:2012ei,Casini:2017vbe} spacetime dimensions. So, in a sense,  the  $a^*_d$ factor in eqn.~(\ref{eq:a-function}) is a generalized central charge,  measuring  the number of degrees of freedom of the~CFT; a  higher dimensional analogue of the central charge $c$ in $d=2$.

In seemingly unrelated work, there has been a great deal of activity in ``extended'' gravitational thermodynamics, a framework that augments traditional black hole thermodynamics\cite{Bekenstein:1973ur,Bekenstein:1974ax,Hawking:1974sw,Hawking:1976de} by making dynamical\footnote{See {\it e.g.} refs.\cite{Henneaux:1984ji,Teitelboim:1985dp,Henneaux:1989zc,Caldarelli:1999xj,Wang:2006eb,Sekiwa:2006qj,LarranagaRubio:2007ut,Kastor:2009wy}, and ref.\cite{Kubiznak:2016qmn} for a survey of the literature with some applications.}  the cosmological constant $\Lambda$, supplying an effective pressure {\it via} $p=-\Lambda/8\pi G$. This modifies the dictionary that translates black hole quantities into thermodynamic ones\cite{Kastor:2009wy}: The mass $M$ becomes the enthalpy $M=H\equiv U+pV$, (where $U$ is the internal energy) and the First Law of thermodynamics\footnote{Here we are neglecting properties such as charge and rotation, which would contribute with standard extra terms.} is $dH=TdS+Vdp$, with the volume emerging as $p$'s conjugate variable: $V=(\partial H/\partial p)|_S$. Two developments in that area are particularly relevant here. The first is the suggestion that since theories with  fixed negative $\Lambda$  are often  holographically dual to field theories,  dynamically varying $\Lambda$ could be meaningful in that context\cite{Kastor:2009wy,Dolan:2013dga,Johnson:2014yja,Kastor:2014dra}. Ref.\cite{Johnson:2014yja} in particular suggested that varying the pressure could be realized as holographic RG flow\cite{Girardello:1998pd,Distler:1998gb}, known to change the effective cosmological constant and hence the degrees of freedom of the dual field theory. The second suggestion is from that same paper where the concept of a ``holographic heat engine'' was proposed. This is a closed cycle in the $(p,V)$ plane, with net heat flows in and out ($Q_H$ and~$Q_C$), resulting in some mechanical work ($W{=}\int \!p\,dV$) being performed. The suggestion  was that this could represent a tour in the space of field theories, and the engine's efficiency ($\eta\,{=}W/Q_H$), could well characterize  an important aspect of that tour that would have a natural bound set by the efficiency of a Carnot engine, following from the Second Law of thermodynamics.

In this paper we show  precise  connections between entanglement entropy (including  the quantity defined in eqn.~(\ref{eq:a-function})), the renormalization group flow and holographic heat engines, giving a concrete realization of the ideas of ref.\cite{Johnson:2014yja}. A clue to the connections is as follows. The cosmological constant in  $(d+1)$--dimensional gravity (and the traditionally defined equivalent length scale $L$) is identified with the pressure  according to:
\begin{equation}\label{eq:pressure}
p=-\frac{\Lambda}{8\pi G}=
  \left(\frac{d(d-1)}{16\pi G}\right)\frac{1}{L^2}\ .
\end{equation}
One of the difficulties in understanding the  role of a {\it dynamical} pressure variable in the field theory is the fact that it is a  dimensionful quantity that does not seem particularly natural. However,  consider two holographic conformal field theories (CFT$_1$ and CFT$_2$, say) with different values of~$a^*_d$. One way to  compare them is {\it via} the ratio of their central charges, which is: 
\begin{equation}
\frac{a_d^{*(1)}}{a_d^{*(2)}}
  =\left(\frac{L_1}{L_2}\right)^{d-1}=
  \left(\frac{p_2}{p_1}\right)^{\frac{d-1}{2}}\ .
  \end{equation}
Notably, ratios such as this appear very naturally in the extended thermodynamics when considering the efficiency of holographic heat engines, {\it e.g.}, for a class of AdS black holes in the large volume limit \cite{Johnson:2014yja} (usually referred as ``ideal gas" holes\cite{Johnson:2015ekr,Johnson:2015fva}).  In fact, ref.\cite{Rosso:2018acz} showed that for a cycle of general shape, the black hole ideal gas engine efficiency is $\eta_{\rm i.g.}\!=1-\langle p \rangle_C/\langle p \rangle_H$
where~$\langle p \rangle$, the mean value of the pressure,  is evaluated along the lower $(C)$ and upper $(H)$ paths of the cycle. This is an important clue in seeing how the heat engine efficiency  can characterize aspects of a field theory tour. We will see precisely how this arises in the context of entanglement entropy in  section~\ref{sec:heat-engines-2}.

Given that work and heat flows are  central quantities in an engine cycle, it would also be nice to have an understanding of their meaning in the field theory. However, as with pressure, it has been difficult to find precise understanding, partly because the correct context had to be found. In this paper we present a context in which we can make precise connections.  Our work builds on the observation of Casini {\it et.~al.} \cite{Casini:2011kv} that  the entanglement entropy arising from  reducing the ground state of a CFT on a  region bounded by a round sphere (a ball) can be mapped to the thermal entropy of a special hyperbolic black hole (with {\it  fixed} negative cosmological constant). We  study the extended thermodynamics of more general hyperbolic black holes in  AdS  and  allow that special  hyperbolic black hole to  join the dynamics of its more general siblings in extended thermodynamics where the cosmological constant (and hence the dual field theory degrees of freedom) can vary. Processes in the thermodynamics will be interpreted as flows and deformations of field theories. Heat engines using these black holes as a working substance will be natural in this setting, and will yield information that has meaning in field theory terms. We will obtain precise field theory understanding of not just the efficiency of these engines but also of their work and heat flows. 

This paper is organized as follows: We start in section \ref{sec:background} with a review of the connection between the massless hyperbolic black hole and the ground state of a CFT in flat space reduced to a sphere. We do this carefully, since some details of the construction will be essential in our following discussions. Section \ref{sec:extended} introduces for the first time the extended thermodynamics of hyperbolic black holes for arbitrary values of mass and (negative) cosmological constant. In section \ref{sec:Dynamics}, we bring together the results and discussions of sections \ref{sec:background} and \ref{sec:extended} in order to provide a field theory description of the extended dynamics of the hyperbolic black hole. In particular, we show how a given path in the $(p,V)$ plane translates exactly into an RG flow in the field theory. Moving to section \ref{sec:heat-engines-1}, we construct three  useful  engines and discuss  their precise meaning in the field theory context. We finish in section~\ref{sec:Conclusion} with a discussion of our main results and future directions of research.

\section{Background}
\label{sec:background}
In this section we review the connection between the ground state of a CFT in Minkowski reduced to a ball and a massless hyperbolic black hole in asymptotic AdS. This will be done in two steps: first by conformally mapping the ground state on the sphere to a thermal state in a hyperbolic space and then by applying the standard AdS/CFT dictionary. Our presentation will follow closely ref.\cite{Casini:2011kv}. Throughout this discussion the AdS radius $L$ is kept fixed. 

\subsection{Reduced Vacuum as Thermal State: Conformal Map}
\label{sec:conformal}
We start by considering the ground state of a CFT in $\mathbb{R}\times\mathbb{R}^{d-1}$ and reduce it to a ball of radius~$R$. \noindent The density matrix of the reduced state is given by\cite{Casini:2011kv}:
\begin{equation}\label{eq:densitymodular}
\rho_{R}=e^{-K_R}\,,
  \qquad \qquad
  K_R=2\pi\int\limits_{|\vec{x}|\le R}d^{d-1}x\,
  \left(\frac{R^2-|\vec{x}|^2}{2R}\right)T_{00}(\vec{x})+{\rm const.}\ ,
\end{equation}
where the constant is fixed to give the density matrix unit trace. Since it is written as an integral of a point--like operator, the modular Hamiltonian $K_R$ is a local operator and generates a local flow of the algebra of operators in the causal domain of the ball.

The Minkowski metric in spherical coordinates is given by $ds^2=-dt^2+dr^2+r^2d\Omega^2_{d-2},$ 
where $d\Omega^2_{d-2}$ is the line element of a unit sphere. Consider the following change of coordinates:
\begin{equation}\label{eq:4}
\big(t,r\big)=\frac{R}{\cosh(u)+\cosh(\tau/R)}
  \Big(
  \sinh(\tau/R), \sinh(u)  
  \Big)\ ,
\end{equation}
where $\tau \in \mathbb{R}$ and $u \in \mathbb{R}_{+}$. 
It is straightforward to show that the new coordinates $(\tau,u)$ only cover the causal domain of the ball and that the originally flat metric is written as:
\begin{equation}\label{eq:hyperbolic}
ds^2=
  \frac{1}{\left(
  \cosh(u)+\cosh(\tau/R)
  \right)^2}
  \Big(
  -d\tau^2+
  R^2\big(
  du^2+\sinh^2(u)d\Omega^2_{d-2}  
  \big)
  \Big)\ ,
\end{equation}
where between parentheses we recognize the line element of $\mathbb{R}\times \mathbb{H}^{d-1}$, with $\mathbb{H}^{d-1}$ a hyperbolic plane of scale size $R$. Applying a conformal transformation we can remove the overall factor and get a CFT on this hyperbolic background. The causal region of the ball in  flat spacetime was mapped to the entire hyperbolic space, while the complementary region was pushed to infinity by the conformal transformation.

Although the ground state is invariant under this conformal transformation, the reduced state $\rho_R$ is not. Since the modular Hamiltonian $K_R$  generates a local flow inside the ball, one might expect  that the transformed operator  generates some local translation in the hyperbolic background. It can be shown \cite{Casini:2011kv} that the transformed operator generates ordinary time translations in $\tau$, meaning that it is thermal. By examining the period of the CFT correlators in imaginary time, its   temperature is determined to be $T=1/(2\pi R)$. If $U$ is the unitary operator acting on the Hilbert space which implements the conformal transformation, we have:
\begin{equation}
\label{eq:thermal}
\rho_R=e^{-K_R}=U^{\dagger}\left(\frac{e^{-\beta H_{\tau}}}{Z}\right)U\ ,
\end{equation}
where $H_{\tau}$ is the hamiltonian in the hyperbolic space which generates time translations in $\tau$. Since the von Neumann entropy is invariant under a unitary transformation, this means that the entanglement entropy across the sphere $S_{\rm EE}=-{\rm Tr }(\rho_R\log \rho_R)$ is mapped to a thermal entropy on the hyperbolic background.

\subsection{Holographic Correspondence}
\label{sec:holography}
It is here that the AdS/CFT correspondence enters the fray in our discussion. The ground state of the CFT in $d$--dimensional flat space will be dual to pure AdS$_{d+1}$ described by a given set of coordinates (Poincare coordinates). On the other hand, the thermal entropy of the CFT on the spacetime $\mathbb{R}\times\mathbb{H}^{d-1}$ can be cast as the Bekenstein--Hawking entropy of a black hole in an  asymptotically AdS$_{d+1}$ spacetime, presented with a hyperbolic slicing\cite{Emparan:1999pm} so that its boundary is~$\mathbb{R}\times\mathbb{H}^{d-1}$. With a bit of care, the precise kind of black hole needed can be uncovered.

We start by recalling the well--known description of AdS as an embedded hypersurface in~$\mathbb{R}^{2,d}$, arising from the constraint:
\begin{equation}
\label{eq:embedding}
-y_{-1}^2-y_{0}^2+y_1^2+...+y_d^2=-L^2\,, 
  \qquad {\rm in} \qquad
  ds^2=-dy_{-1}^2-dy_0^2+dy_1^2+...+dy_d^2\ .
\end{equation}
One description of this surface comes from the choice of Poincare coordinates $(z,x^a)$ :
\begin{equation}
\label{eq:10}
y_{-1}+y_d=\frac{L^2}{z}\,,
  \qquad \qquad
  y_a=\frac{L}{z}x^a\ ,
\end{equation}
($a=0,\hdots,d-1$), so that the  constraint and metric (\ref{eq:embedding}) become:
\begin{equation}
y_{-1}-y_d=z+\frac{1}{z}\eta_{ab}x^ax^b
  \qquad {\rm and} \qquad
  ds^2=\left(\frac{L}{z}\right)^2
  \left(
  dz^2+\eta_{ab}dx^adx^b
  \right)\ ,
  \end{equation}
where the boundary is at  $z\rightarrow 0$.  Taking this limit and removing the conformal factor we recover Minkowski spacetime $\mathbb{R}\times \mathbb{R}^{d-1}$. Another set of coordinates to describe pure AdS$_{d+1}$ is:
\begin{eqnarray}
y_{-1}&=&\cosh(\gamma)\rho \cosh(u)+\sinh(\gamma)\rho'\cosh(\tau'/L)\,,  \nonumber
 \\
y_d&=&\sinh(\gamma)\rho \cosh(u)+\cosh(\gamma)\rho'\cosh(\tau'/L)\ ,  \nonumber
\\
y_0&=&\rho'\sinh(\tau'/L)\,,
  \quad
  y_1=\rho \sinh(u)\cos(\phi_1)\,,
  \nonumber \\
y_2&=&\rho \sinh(u)\sin(\phi_1)\cos(\phi_2)
  \quad  \cdots  \quad 
  y_{d-1}=\rho \sinh(u)\cdots\sin(\phi_{d-2})\ ,
 \label{eq:12}
\end{eqnarray}
where  $\gamma \in \mathbb{R}$ is not  a coordinate but  a fixed boost parameter (acting on the coordinates $y_{-1}$~and~$y_d$), as can be seen from the form of the first two lines. 
The constraint that defines the surface in eq.~(\ref{eq:embedding}) becomes $\rho'=\sqrt{\rho^2-L^2}$, so that the resulting metric  is given by:
\begin{equation}\label{eq:special-hyperbolic}
ds^2=-\left(\frac{\rho^2}{L^2}-1\right)d\tau'^2+
  \frac{d\rho^2}{\left(\frac{\rho^2}{L^2}-1\right)}+\rho^2\left(du^2+\sinh^2(u)d\Omega_{d-2}^2\right)\ ,
\end{equation}
with $\rho \in [L,+\infty)$ and $u\in \mathbb{R}_{+}$.  Despite  the fact that the boost parameter $\gamma$ does not appear on the metric, it will play a crucial role in the following discussion.

The metric (\ref{eq:special-hyperbolic}) can be interpreted as a black hole in AdS$_{d+1}$ with a hyperbolic horizon, temperature $T_{\rm BH}=1/(2\pi L)$ and vanishing mass. It is a special case of a more general class of hyperbolic black holes we will explore in section~\ref{sec:extended}. 

To show that metric~(\ref{eq:special-hyperbolic}) describes the thermal state obtained in the previous section, we must first find  the coordinate transformation that relates the Poincare (\ref{eq:10}) and hyperbolic (\ref{eq:12}) bulk coordinates on the boundary. A straightforward calculation yields the following
\begin{equation}
\big(t,r\big)=
  \frac{e^{-\gamma}L}{\cosh(u)+\cosh(\tau'/L)}
  \Big(\sinh(\tau'/L),\sinh(u)\Big) \ ,
  \end{equation}
where  $r^2=x_1^2+...+x_{d-1}^2$, and $t$ and $r$ are the time and spatial radial coordinates of Minkowski spacetime. This is exactly the same coordinate transformation we previously applied (see eq.~(\ref{eq:4})) to map the ball in Minkowski to $\mathbb{R}\times\mathbb{H}^{d-1}$,  if we make the identifications: 
\begin{equation}\label{eq:identify}
R=e^{-\gamma}L\,,
  \qquad \qquad
  \tau'=e^{\gamma}\tau
  \ .
\end{equation}
This means that changing between these set of coordinates in the bulk is equivalent to the discussion we had in the previous section in the boundary theory. Using eq.~(\ref{eq:identify}), with a choice of $\gamma$, we can now take the asymptotic limit $\rho \rightarrow +\infty$ in eq.~(\ref{eq:special-hyperbolic}) and recover eq.~(\ref{eq:hyperbolic}) up to a conformal factor.  We then conclude that the  spacetime~(\ref{eq:special-hyperbolic})  together with the identifications in eq.~(\ref{eq:identify}),  gives the holographic description of the thermal state described in section \ref{sec:conformal}, and therefore of the ground state of the CFT in flat space reduced to a ball.


\subsection{Entanglement Entropy from Horizon Area}
\label{sec:entropy}
We can now calculate the entanglement entropy of the ground state reduced to a ball in Minkowski from the black hole entropy of eq.~(\ref{eq:special-hyperbolic}). Using standard methods, its temperature and entropy are given by\cite{Emparan:1999pm,Emparan:1999gf}:
\begin{equation}\label{eq:temp-ent}
T_{\rm BH}^{(M=0)}=\frac{1}{2\pi L}\ ,
  \qquad \quad
  S_{\rm BH}^{(M=0)}=\frac{w_{d-1}L^{d-1}}{4G}=\frac{L^{d-1}}{4G}
 \Omega_{d-2}
  \int_{0}^{+\infty} du\,
  \sinh^{d-2}(u)\ ,
\end{equation}
where $w_{d-1}$ is the volume ({\it i.e.,} surface area) of the hyperbolic plane with radius one and in the final term we have written it more explicitly, with $\Omega_{d-2}$ the volume ({\it i.e.,} surface area) of a  unit sphere $S^{d-2}$. The horizon  is the hyperbolic space $\mathbb{H}^{d-1}$ which is of infinite extent, a fact captured by  the divergence of the $u$ integral and in accordance with the divergent nature of the entanglement entropy. We must then introduce a cutoff $u_{\rm max}$ which is a cutoff~$z_{\rm min}$ on the Poincare coordinate $z$. From the definition of the Poincare and hyperbolic coordinates (\ref{eq:10}) and (\ref{eq:12}), the black hole horizon $\rho=L$ can be written in as:
\begin{equation}
y_{-1}+y_{d}=\frac{L^2}{z}=e^{\gamma} L \cosh(u)\,, \qquad \qquad
  y_1^2+...+y_{d-1}^2=\frac{L^2}{z^2}r^2=L^2\sinh^2(u)\ ,\end{equation}
  giving
\begin{equation}\label{eq:relations}
\left(z,r\right)=e^{-\gamma} L
  \left(\frac{1}{\cosh(u)},|\tanh(u)|\right)\ .
\end{equation}
Taking the limit $u\rightarrow \infty$ shows that $(z,r)\rightarrow (0,R)$, so that we can write the  cutoff as:
\begin{equation}
\label{eq:cutoff}
z_{\rm min}=\frac{R}{\cosh(u_{\rm max})}
  \qquad \Longrightarrow \qquad
  x_{\rm max}=\sinh(u_{\rm max})=\sqrt{\left(\frac{R}{\epsilon}\right)^2-1}\ ,\end{equation}
where we have defined $z_{\rm min}=\epsilon$. Changing the integration variable in eq.~(\ref{eq:temp-ent}) to $x=\sinh(u)$, the regulated entropy is given by:
\begin{equation}
\label{eq:entanglement}
S_{\rm EE}=
  \left(
  \frac{2\Gamma(d/2)\Omega_{d-2}}{\pi^{d/2-1}}
  \right)
  a^*_d
  \int_{0}^{x_{\rm max}} dx\,
  \frac{x^{d-2}}{\sqrt{1+x^2}}\,,
  \quad{\rm with}\quad T_{\rm BH}^{(M=0)}\,\,\Longrightarrow \,\,T=\frac{1}{2\pi R}\ ,
\end{equation}
where the $L$ dependent factor was written in terms of $a^*_d$, the generalized central charge, from eq.~(\ref{eq:a-function}),  and we also used eq.~(\ref{eq:identify}) to recover $T$, the  temperature appearing in eq.~(\ref{eq:thermal}). The integral in the entanglement entropy can be evaluated explicitly and written in terms of a hypergeometric function.  Notice that all the information about the radius $R$ of the ball comes from solving this integral and is contained in the cutoff $x_{\rm max}$. It is through differing values of $a^*_d$ that differences between CFTs can make their presence felt.

\section{Hyperbolic Black Holes in Extended Thermodynamics}
\label{sec:extended}
In this section we consider the generalization of the metric~(\ref{eq:special-hyperbolic}) to arbitrary values of mass, in Einstein's gravity with negative cosmological constant. Although many standard thermodynamic properties of these black holes in arbitrary dimensions were previously studied in refs.~\cite{Emparan:1999pm,Emparan:1999gf,Cai:2004pz}, we will present some of their richer \textit{extended} thermodynamics here for the first time. The metric is given by\cite{Birmingham:1998nr}\footnote{Four dimensional versions of these hyperbolic black holes first appeared in the literature as ``topological black holes'' in refs.\cite{Mann:1996gj,Vanzo:1997gw,Brill:1997mf,Emparan:1998he}, accompanied by some discussion of their thermodynamics.}:
\begin{equation}
ds^2=-V(\rho)d\tau'^2+\frac{d\rho^2}{V(\rho)}+\rho^2\left(du^2+\sinh^2(u)d\Omega^2_{d-2}\right)\ ,
\end{equation}
where
\begin{equation}
V(\rho)=\frac{\rho^2}{L^2}-\frac{\mu}{\rho^{d-2}}-1\ .
\end{equation}
The horizon radius $\rho=\rho_+$ is defined by the vanishing  $V(\rho_+)=0$, while the factor $\mu$ is proportional to the (ADM) mass of the black hole:
\begin{equation}\label{eq:mass}
M(\rho_+,L)=\left(\frac{(d-1)w_{d-1}}{16\pi G}\right)\mu=\left(\frac{(d-1)w_{d-1}}{16\pi G}\right)
  \rho_+^{d-2}\left[\left(\frac{\rho_+}{L}\right)^2-1\right]\,,
\end{equation}
where $w_{d-1}$ is the volume of the hyperbolic space with radius one (whose regularization was discussed in section~\ref{sec:entropy}). In the extended thermodynamics, the mass of a black hole is identified with the enthalpy\cite{Kastor:2009wy}: $M=H(S,p)=U+pV$, and the First Law can be written as $dH=TdS+Vdp$. We should then write eq.~(\ref{eq:mass}) explicitly in terms of the pressure (using its relation in eq.~(\ref{eq:pressure}) to the AdS radius), and the entropy, given by the Bekenstein--Hawking law:
\begin{equation}\label{eq:26}
S=\frac{w_{d-1}\rho_+^{d-1}}{4G}\ .
\end{equation}
The black hole volume and temperature are readily computed to be:
\begin{eqnarray}\label{eq:the-volume}
V&=&\left.\frac{\partial M}{\partial p}\right|_{S}=
  \left.\frac{\partial M}{\partial p}\right|_{\rho_+}=
  \frac{w_{d-1}\rho_+^{d}}{d}=\left(\frac{4G}{d}\right)\rho_+S\ ,
\\
\label{eq:the-temperature}
T&=&\left.\frac{\partial M}{\partial S}\right|_p=
  \left.\frac{\partial M }{\partial \rho_+}\right|_{p}
  \left.\frac{\partial \rho_+ }{\partial S}\right|_{p}=
  \frac{(d-2)}{4\pi}\frac{1}{\rho_+}
  \left[\frac{16\pi G }{(d-1)(d-2)}\rho_+^2p-1\right]\ .
\end{eqnarray}
Notice that the volume is given by the naive geometrical volume, which only depends in $\rho_+$. Since the entropy (\ref{eq:26}) is also only a function of $\rho_+$, we conclude that isochoric paths (constant $V$) are equal to adiabatics (constant $S$) in the whole $(p,V)$ plane \cite{Johnson:2014yja}.

This hyperbolic black hole has some unique features, as can be seen by calculating the zero mass and zero temperature curves in the $(p,V)$ plane:
\begin{equation}\label{eq:mapping}
p(M=0)=\frac{\kappa}{V^{2/d}}\ ,
  \qquad \qquad
  p(T=0)=\left(\frac{d-2}{d}\right)\frac{\kappa}{V^{2/d}}\ ,
\end{equation}
where
\begin{equation}
\kappa=\frac{d(d-1)}{16\pi G}\left(\frac{w_{d-1}}{d}\right)^{2/d}\ .
\end{equation}
From eq.~(\ref{eq:mass}), notice that the zero mass curve can also be written as the condition $\rho_+=L$.
 \begin{wrapfigure}{l}{0.4\textwidth}
 {
\centering
\includegraphics[width=0.35\textwidth]{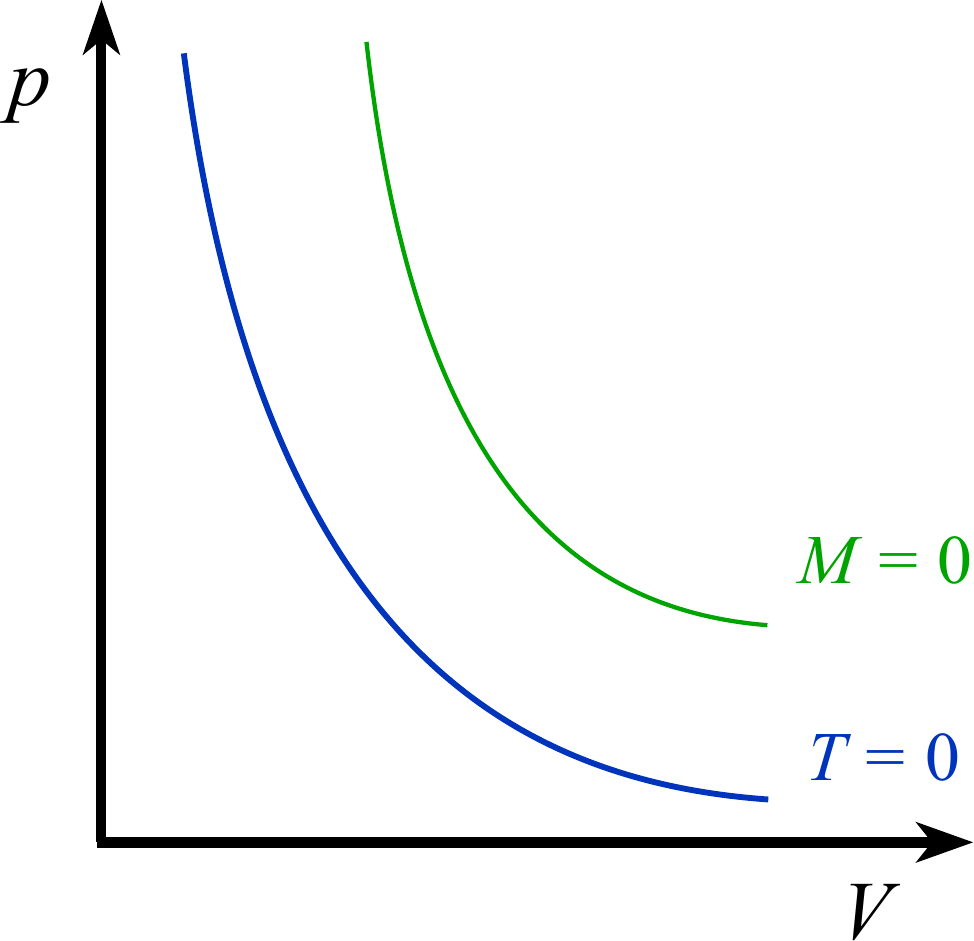}
\caption{\label{fig:NegativeMass}The upper  curve describes $M{=}0$ black holes, while the lower  curve  $T{=}0$ holes. The  region between them (including the lower curve) are black holes with $M{<}0$.}
}
\end{wrapfigure}
The first  remarkable feature
is the fact these curves  do not coincide.
 In fact, there is a whole region in the $(p,V)$ plane below the $M=0$ curve where the mass of the black hole is negative (see figure \ref{fig:NegativeMass}). Although this curious behaviour is  known in the literature\cite{Emparan:1999pm,Emparan:1999gf}, it becomes richer in the extended thermodynamics and it will have its consequences when applying the mapping from the previous section. In section~\ref{sec:FirstLaw} we discuss its meaning in the dual field theory.

Before moving on and establishing the connection with the field theory, it will be useful to calculate the work and heat flow for a process that moves along the zero mass curve. The work is given by:
\begin{equation}\label{eq:work}
W^{(M=0)}=
  \int_{V_i}^{V_f}p\,dV=
  \left(\frac{d}{d-2}\right)\Delta\left(pV\right)\ ,
  \end{equation}
where we have considered $d\ge 3$ and used eq.~(\ref{eq:mapping}). Since the mass of the black hole vanishes along this path, we can calculate the heat flow from
$0=\Delta M=\Delta U+\Delta\left(pV\right)=
  Q-W+\Delta\left(pV\right),$
which gives:
\begin{equation}
\label{eq:heat}
  Q^{(M=0)}=\left(\frac{2}{d-2}\right)\Delta\left(pV\right)\ .
\end{equation}
Notice that both the work and heat flow along this path are proportional to the change in $pV$. We can use Smarr's relation \cite{Smarr:1972kt,Kastor:2009wy} to express them in terms of different thermodynamic variables:
\begin{equation}\label{eq:Smarr}
\left(\frac{d-2}{d-1}\right)M=
  TS-\left(\frac{2}{d-1}\right)pV
  \qquad \Longrightarrow \qquad
  pV=\left(\frac{d-1}{2}\right)TS\ ,
\end{equation}
where we have used that we are on the massless curve. Using this in (\ref{eq:work}) and (\ref{eq:heat}) we find:
\begin{equation}\label{eq:work-heat}
W^{(M=0)}=\frac{d}{2}
  \left(\frac{d-1}{d-2}\right)
  \Delta\left(TS\right)\,,
  \qquad \qquad
  Q^{(M=0)}=
  \left(\frac{d-1}{d-2}\right)
  \Delta\left(TS\right)\ .
\end{equation}
This rewriting of the work and heat in terms of the entropy and temperature will be a key point in our upcoming discussions.

\section{CFT Description of Black Hole Dynamics}
\label{sec:Dynamics}
In this section we bring together the discussions of sections \ref{sec:background} and \ref{sec:extended} in order to illustrate the field theory meaning of the hyperbolic black hole dynamics. First we show that moving toward higher~$p$ along the massless curve of the black hole (\ref{eq:mapping}) is {\it exactly} equivalent to implementing an RG flow in the field theory. We then comment on the field theory meaning of leaving the zero mass curve and its relation to the First Law of entanglement \cite{Blanco:2013joa}.  We also show that black holes with small negative values of mass describe perturbations from the ground state with negative energy density inside the ball.

\subsection{RG Flow and Massless Black Holes}
\label{sec:RGFlow}
\begin{wrapfigure}{l}{0.4\textwidth}
\centering
\includegraphics[width=0.35\textwidth]{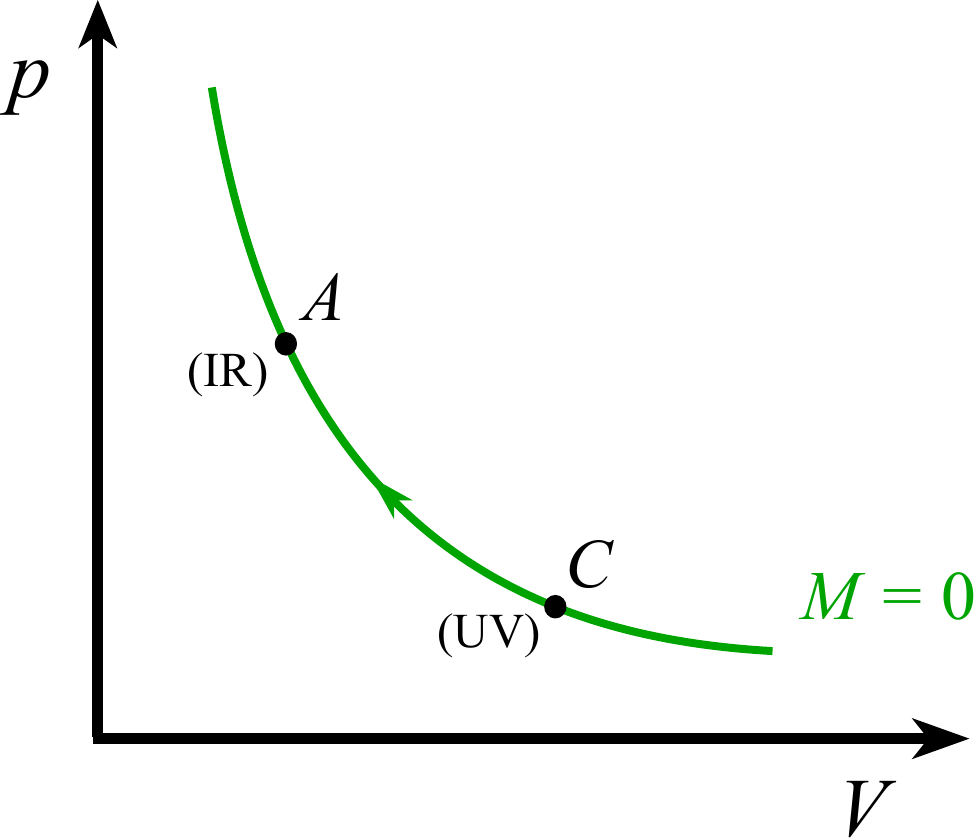}
\caption{\label{fig:HoloRG} Two different massless black holes at points $C$ and $A$ connected by the zero mass curve. Since $p\propto 1/L^2$ and $a^*_d\propto L^{d-1}$, point~$C$ describes  a CFT$_C$ with a larger number of degrees of freedom than CFT$_A$.}
\end{wrapfigure}

Let's start by considering a random point (say $C$) on the massless curve. From the discussion in section \ref{sec:background}, we know that this black hole will be dual to the ground state of a CFT in flat space, 
reduced to a spherical region of  radius we will call $R_C$, which, through eq.~(\ref{eq:identify}) is set by $R_C=e^{-\gamma}L_C$. This radius is what we use in eqs.~(\ref{eq:cutoff}) and (\ref{eq:entanglement}) to evaluate the entanglement entropy and associated  temperature. Now let's move along this special curve towards a point at higher pressure, $A$, as shown in fig.~\ref{fig:HoloRG}.  Since higher $p$ means lower~$L$, the central charge $a^*_d$  changes according to (see eqs.~(\ref{eq:a-function})  and~(\ref{eq:pressure})):
\begin{equation}\label{eq:new-dof}
a_d^{*(C)} \quad \longrightarrow \quad a_d^{*(A)}=
b^{d-1}
a_d^{*(C)}\ ,
\end{equation}
where $b\equiv L_A/L_C<1$. This means that the number of degrees of freedom of the field theory decreases along the path. What about the radius of the region the state is reduced on in theory~$A$? 
Since we have fixed $\gamma$,  from eq.~(\ref{eq:identify}) we have that the radius is
$  R_A=bR_C$,
meaning that  the ball in the CFT$_A$ is smaller. This  is exactly what we expect from an RG flow: By going deeper into the infrared (IR), we are focusing on  longer wavelength physics, and so a (fixed) length scale (such as  the radius of the ball) effectively looks {\it shorter}.

There is an alternative (and useful) way of  interpreting the  physics, by thinking in terms of  the explicit  cutoff $\epsilon$ in eq.~(\ref{eq:cutoff}), which should have {\it increased} by flowing into the IR. We can read this off by looking how eq.~(\ref{eq:cutoff}) changes when going from $C$ to $A$:
\begin{equation}
\label{new_cutoff}
x_{\rm max}^{(C)} \,\, \longrightarrow \,\, x_{\rm max}^{(A)}=
  \sqrt{\left(\frac{R_A}{\epsilon}\right)^2-1}=
  \sqrt{\left(\frac{R_C}{\epsilon/b}\right)^2-1}\ .
\end{equation}
As expected, we obtain a larger cutoff $\epsilon^\prime=\epsilon/b$, whereas the radius of the ball at $A$ is unchanged and still given by $R_C$.\footnote{Naively, in this picture where the radius stays fixed at $R_C$, it looks like point~$A$ has   temperature $1/(2\pi R_C)$, but that is not the case as we have seen from the other picture: Deeper into the IR the effective temperature is larger by a factor $b^{-1}$: $T_A=1/(2\pi R_A)$.}
 Regardless of how one looks at it, CFT$_A$ has a  smaller  number of degrees of freedom than CFT$_C$ (see eq.~(\ref{eq:new-dof})).

When interpreting this process we must keep in mind that our basic framework is equilibrium thermodynamics, which has certain tacit assumptions that should be recalled. Any  point in the $(p,V)$ plane (including ones on the  massless curve) corresponds to  an equilibrium state. Moving to another point ({\it e.g.,} along the massless curve) requires that  we perturb the system. In other words,  intermediate states are non--equilibrium states. But for standard equilibrium thermodynamics a process such as the one in figure \ref{fig:HoloRG} is usually considered as taking place in the \textit{quasi--static} limit, in which the perturbation is infinitesimally small and resulting changes are slow enough for us to forego a description of those non--equilibrium states, exhibiting instead a sequence of equilibrium points. However, we must not forget that in principle such states are always present in order to give meaning to moving around the plane.

How should we interpret this from the field theory perspective? An equilibrium point in the massless curves corresponds to the ground state of a CFT reduced to a ball, while the non-equilibrium states correspond to field theory  deformations away from the CFT. This means that although the process of going from $C$ to $A$ described in figure \ref{fig:HoloRG} seems to continuously connect CFTs (perhaps calling into question its identification as an RG flow\footnote{We thank an anonymous referee for a helpful question that prompted us to clarify this point.}),  this is certainly not the case, due to the presence of intermediate non-equilibrium states (CFT deformations). It is an artefact of working with the equilibrium thermodynamics description.

A related fact is that  the extended thermodynamics cannot say much (if anything)  about the detailed operator description of how the RG flow described in figure \ref{fig:HoloRG} is triggered.  Consider the standard thermodynamics of a gas. It is a very powerful framework that does not rely on the specific microscopic details and motions  of the gas, dealing only with macroscopic quantities (such as $p$, $V$, and $T$)  and giving useful answers in terms of such variables. In the same way, the extended thermodynamics we are using here captures useful (and sometimes universal) information about the field theory RG flows,  but can shed little light on details more appropriate in a Lagrangian field theory description.   Nevertheless, we can learn a lot from this thermodynamic connection.

Going from the ultraviolet (UV) to the IR by integrating out and discarding  short distance details should be an irreversible process. What is the meaning of this in the extended thermodynamics setup? In this context, going from $C$ to $A$ corresponds to a process in which work is done on the system, with some resulting outflow of heat\footnote{Microscopically, coarse--graining reduces the number of available configurations available to the system, so $\dbar Q=TdS<0$. Note that $Q$ and $W$ can be computed from our results in equation~({\ref{eq:work-heat}}), where they have the same sign, negative in this case.}. Crucially, the outward heat flow requires contact with a  reservoir which is at a lower temperature  than any point on the $C$--$A$ curve.  In this picture then, the irreversibility of flowing from $C$ to $A$ that we expect in a true field theory RG flow simply follows from the Second Law of thermodynamics in this extended setting: Trying to reverse the flow to head back in the the opposite direction would have heat flowing from the low temperature reservoir to the higher temperatures that lie between $A$ and $C$. This is forbidden. So we have the pleasing result that the properties of an RG flow in a field theory are not just analogous to those of a thermodynamics heat flow, but {\it exactly equivalent} in  this framework.

RG flows such as this one have been previously studied and some also have holographic RG descriptions that are well known. An example in $d=4$ is the flow from ${\cal N}=4$ supersymmetric $SU(N)$ Yang--Mills (in the  large $N$ limit) to an  ${\cal N}=1$ superconformal fixed point by turning on a mass for one of the three ${\cal N}=1$ chiral multiplets that makes up the ${\cal N}=4$ vector multiplet. The gauge theory  and the holographic supergravity solution are described in refs.\cite{Leigh:1995ep,Karch:1999pv,Khavaev:1998fb,Freedman:1999gp,Pilch:2000fu}. In this example the value of the scaling parameter $b$ can be exactly calculated and is given by $b=3/2^{\frac{5}{3}}$.  In the gauged supergravity description the flow connects two AdS vacua, and  their cosmological constants $\Lambda_{\rm UV/IR}$  are set by the values of the scalars in the theory. The values of the scalars change as we move from one vacuum to another, resulting in $|\Lambda_{\rm IR}|{>}|\Lambda_{\rm UV}|$. This is captured by our extended thermodynamics flow from $C$ to $A$ along the $M=0$ curve (where we repeat the caveat mentioned a few paragraphs above that moving along the curve necessarily involves deformations of CFT).

\subsection{First Law of Entanglement and Massive Black Holes}
\label{sec:FirstLaw}

We now examine what happens to the field theory as we move away from the massless curve. At least infinitesimally, we will be able to give meaning to such motions by considering the (extended) First Law of black hole thermodynamics (remember that $M=H\equiv U+pV$, the enthalpy):
\begin{equation}
\label{eq:First-Law}
dM=TdS+Vdp\ .
\end{equation}
We begin on the $M=0$ curve, with $T$ and $V$  evaluated on it. Using Smarr's relation~(\ref{eq:Smarr}), and writing the pressure in terms of $a^*_d$ (see eq.~(\ref{eq:a-function})) we find:
\begin{equation}\label{eq:First-Law-2}
\frac{dM}{T}=dS+\left(\frac{d-1}{2}\right)S \frac{dp}{p}=
  dS-\frac{S}{a^*_d} da^*_d\ .
\end{equation}
All of the quantities in this equation have a CFT interpretation. For a given CFT (which implies a fixed value of $a^*_d$) it was argued in ref.\cite{Faulkner:2013ica} that $dS$ should be translated as $\delta S_{\rm EE}$, the difference between  the entanglement entropy of the system in the ground state and that of a perturbation, both reduced to a ball of radius $R$.  On the other hand, the mass on the left hand side is given by the conserved charge associated to the Killing vector $\partial_{\tau'}$. Writing  this Killing vector in terms of the Poincare coordinates $(z,t,\vec{x})$, using (\ref{eq:identify}) and considering the boundary limit $z\rightarrow 0$ yields\cite{Faulkner:2013ica}:
\begin{equation}\label{eq:29}
\frac{dM}{T}=2\pi \!\! \int\limits_{|\vec{x}|\le R} d^{d-1}x\,\left(\frac{R^2-|\vec{x}|^2}{2R}\right)\delta
  \langle T_{00}(\vec{x}) \rangle=
  \delta \langle K_R \rangle\ ,
  \end{equation}
where $\delta \langle \,\cdots \rangle$ means the expectation value of the perturbation from the ground state. We recognize the operator as the modular Hamiltonian of the ground state reduced to a ball (see eq.~(\ref{eq:densitymodular})). So the field theory version of eq.~(\ref{eq:First-Law-2}) is given by:
\begin{equation}\label{eq:First-Law-Ent}
\delta \langle K_R \rangle=\delta S_{\rm EE}-\frac{S_{\rm EE}}{a^*_d}\delta a^*_d\ .
\end{equation}
This is the extended First Law of entanglement \cite{Blanco:2013joa}. The derivation of (\ref{eq:First-Law-Ent}) from the First Law of black hole thermodynamics was previously presented in ref.~\cite{Faulkner:2013ica} (and its extension in ref.~\cite{Kastor:2014dra}).

Our proposal in this paper is that the physics of the points visited by such perturbations is governed by the full family of hyperbolic black holes in  the extended thermodynamics unpacked earlier in section~\ref{sec:extended}. The previous works\cite{Blanco:2013joa,Faulkner:2013ica,Kastor:2014dra} were equivalent to  considering any point on, and small perturbations around, the zero mass curve in the $(p,V)$ plane\footnote{After an earlier version of this manuscript appeared, we learned of ref.~\cite{Hung:2011nu}, which shows how to use the $M\neq0$ hyperbolic black holes (for {\it fixed}  $L$)  to compute the R\'{e}nyi entropy of the ground state reduced to a ball. Their technique has since been applied and developed further in the literature. Note that this use of the hyperbolic black holes is {\it complementary} to the applications we develop for them here.}. 
 
 \begin{wrapfigure}{l}{0.4\textwidth}
{
\centering
\includegraphics[width=0.35\textwidth]{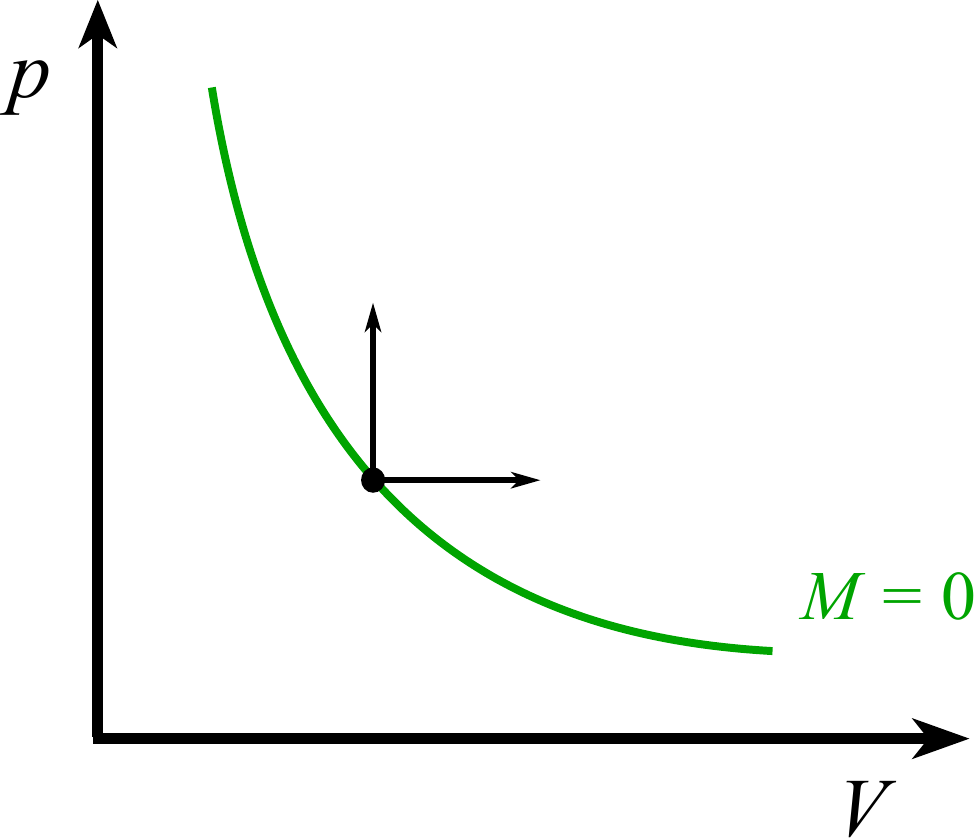}
\caption{\label{fig:OutCurve} Two orthogonal perturbations away from the $M{=}0$  curve.}
}
\end{wrapfigure}
A perturbation away from the massless curve at fixed pressure (a horizontal move in figure \ref{fig:OutCurve}) is equivalent to perturbing the ground state of a fixed CFT, without changing the radius~$R$ of the reduction ball. Meanwhile, perturbing at fixed volume (vertical move in fig.~\ref{fig:OutCurve}) means that we  move towards a different CFT with different central charge $a_d^{*}$, while keeping the entanglement entropy and radius $R$ constant. Either of these moves results in a change $\delta \langle T_{00} \rangle\neq 0$, equivalent to changing the mass on the gravity side, {\it i.e.,} moving off the $M=0$ curve to more general hyperbolic black holes. Staying on the massless curve can then be seen as a special combination of these basic moves that keeps $dM{=}0$. 

In the previous section we showed an exact equivalence between  field theory RG flows and thermodynamic processes obtained by moving up the massless curve. Such flows have known holographic duals as supergravity solutions.  It is natural to conjecture that there are other pressure--increasing moves in the $(p,V)$ plane (away from the massless curve) that correspond to  known holographic RG flows. It would be instructive to explore this in future work.

It is interesting to consider the field theory meaning of the negative mass of the black hole, at least for infinitesimal points below the massless curve (see figure \ref{fig:NegativeMass}). Moving below this curve means $dM<0$ which from (\ref{eq:29}), implies $\delta \langle K_R\rangle<0$. From the integral expression of $K_R$ we see that this necessarily means that the perturbed state has some negative energy density inside the ball $\delta\langle T_{00}(\vec{x}) \rangle<0$. The fact that quantum field theories have states with negative energy density has been known for a long time \cite{Epstein:1965zza} and thoroughly studied in the literature (see ref.~\cite{Fewster:2012yh} for a review). From the First Law of entanglement (\ref{eq:First-Law-Ent}), we can in particular consider a horizontal move towards the negative mass region, so that we find:
\begin{equation}
S_{\rm EE}(\delta \rho_R)-S_{\rm EE}(\rho_R)=
  \delta S_{\rm EE}=
  \delta\langle K_R \rangle<0\ ,
  \quad {\rm yielding:} \quad
  S_{\rm EE}(\delta \rho_R)<S_{\rm EE}(\rho_R)\ ,
  \end{equation}
which means that the perturbed state is less entangled than the original ground state. This means that a state which has negative energy density becomes less entangled, a field theory feature that has been previously studied (see refs.~\cite{Blanco:2013lea,Bianchi:2014qua,Blanco:2017akw}). Our proposal then suggests that small negative values of mass of the hyperbolic black hole describe perturbations from the ground state in the conformal field theory with negative energy density  inside the ball and which are less entangled than the unperturbed state. In this light, it would be interesting to further study the properties of these black holes more.

Since we have used the  First Law to explore infinitesimal perturbations away from the $M=0$ curve, a field theory interpretation for  points in the $(p,V)$ plane  arbitrarily far away from the massless curve is more tentative. The simplest suggestion is  that moving a finite horizontal distance from $M=0$ (changing $V$)  remains equivalent to exploring the Hilbert space of the starting  CFT at $M=0$, while moving a finite vertical distance (changing $p$)  remains equivalent to  not only changing the CFT itself but also the state.

\section{Holographic Heat Engines}
\label{sec:heat-engines-1}
Heat engines in the extended thermodynamics were proposed in ref.\cite{Johnson:2014yja}, where it was conjectured that for negative cosmological constant they would have some interpretation in field theory. While qualitatively suggestive, it was difficult to make a clear connection that could be computationally verified since, in general, the finite temperature nature of the construction required knowledge about  finite temperature holographic flows, which are hard to construct. Here, we have a setting that sidesteps this problem, since (at least near   the massless curve), there is non--zero  temperature on the gravity side, but the field theory of interest is at zero temperature. We will construct closed cycles in the $(p,V)$ plane  and (if moving clockwise) see them live up to their name as  ``holographic heat engines''\footnote{Moving anti--clockwise is possible too, in which case they would be ``holographic refrigerators".}. As we shall see, the key quantity that characterizes a cycle, the efficiency $\eta$, will be able to be written entirely in terms of quantities that have meaning in the field theory. While there may be more specific meanings to~$\eta$ to be discovered than we will explore in this paper, being able to write it in terms of field theory quantities is already a significant step. We will also be able to give complete field theory meaning to the work and heat flows of  some engines. 

We start in section \ref{sec:Special} by constructing some useful heat engines and calculating their thermodynamic quantities. This discussion is done entirely in the context of extended thermodynamics. In section \ref{sec:heat-engines-2} we translate the heat engine thermodynamics into field theory quantities and discuss their meaning.

\subsection{Some Useful Engines}
\label{sec:Special}
\begin{figure}[h]
\centering
\subfigure[]{\centering\includegraphics[scale=0.57]{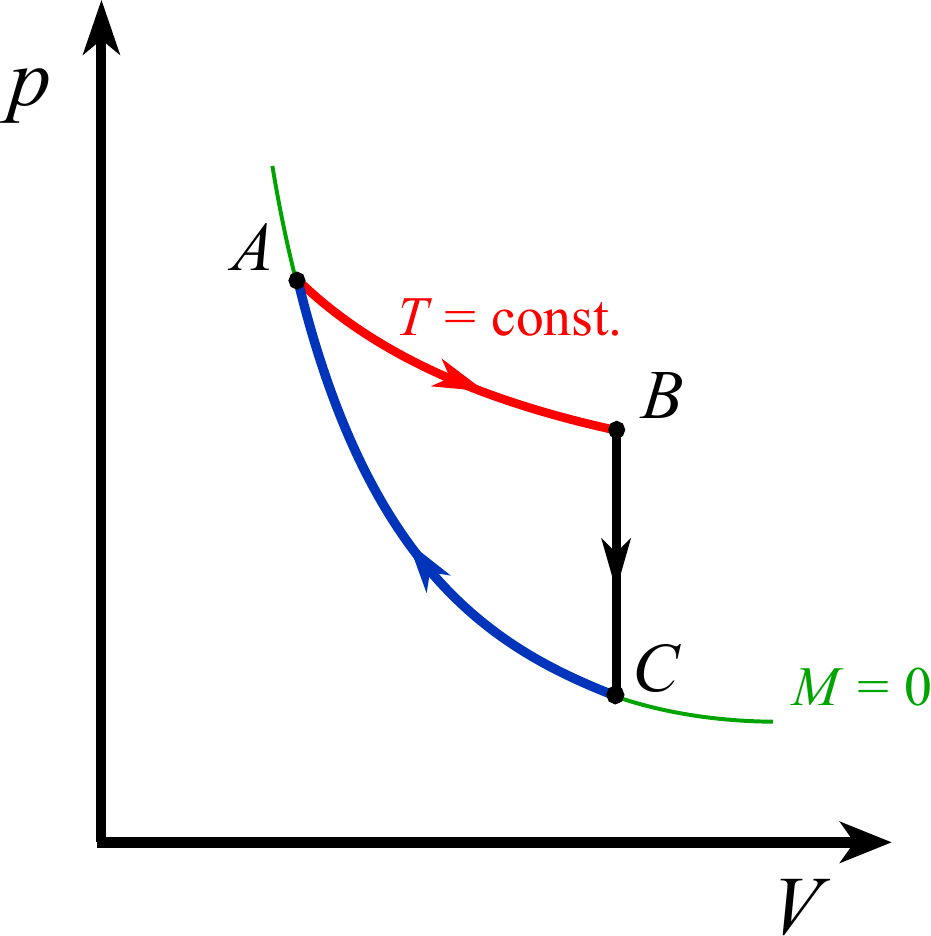}
\label{fig:engine-one-upper}}
\hspace{50pt}
\subfigure[]{\centering\includegraphics[scale=0.57]{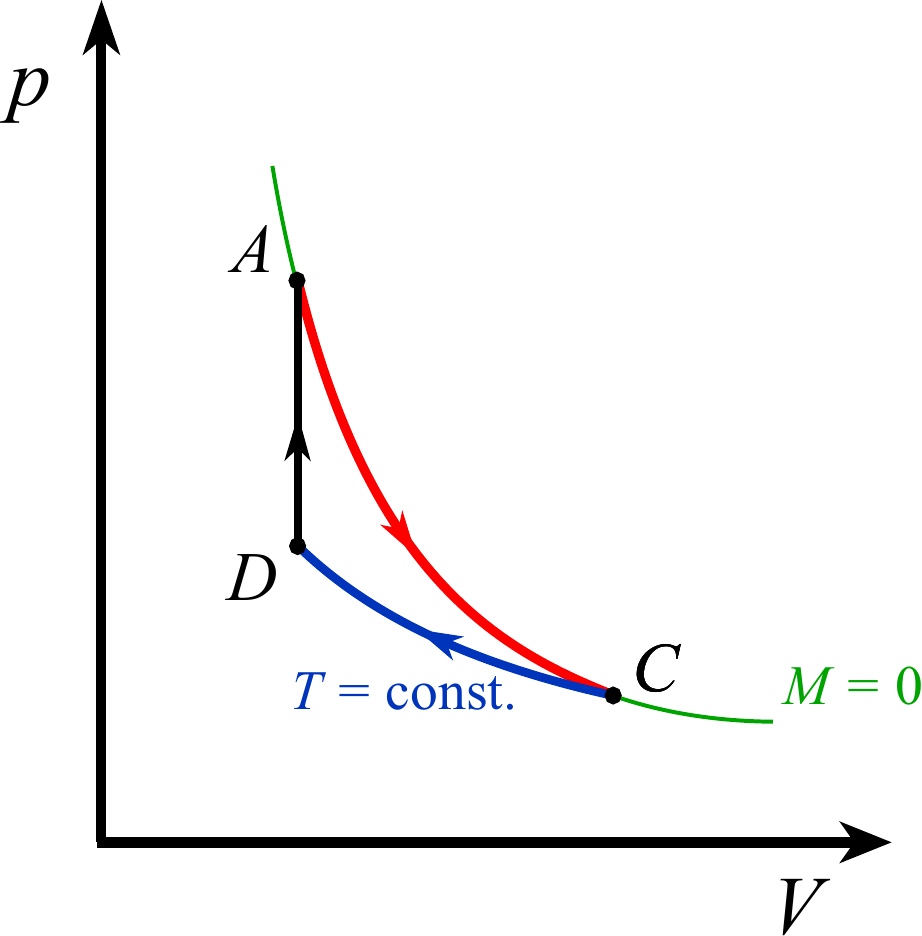}
\label{fig:engine-one-lower}
}
\caption{Two heat engines. In green we see the massless curve given in eqn.~(\ref{eq:mapping}). { (a)} The red path from $A$ to $B$ is an isotherm, and the black path from $B$ to $C$ an isochore/adiabat. { (b)} The blue path from $C$ to $D$ is an isotherm, and the black path from $D$ to $A$ an isochore/adiabat.}\label{fig:engine}
\end{figure}

Consider the path from $C$ to $A$ in fig.~\ref{fig:HoloRG}, which can be embedded into a heat engine cycle by moving along an isotherm from $A$ to point $B$ and then down an isochore/adiabat\footnote{See the discussion of section \ref{sec:extended}: Isochoric paths are equal to adiabatic ones because the entropy and volume are not independent variables. This was noticed as a feature for all static black holes in ref.\cite{Johnson:2014yja}.} back to $C$. This cycle can be seen in figure \ref{fig:engine-one-upper}. The reason for constructing an engine in this way will become clear in a moment.
\noindent The engine's efficiency can be written in terms of the heat flows as
$\eta=1-{Q_C}/{Q_H}$,
where~$Q_C$ is the magnitude of the heat flowing out. As noticed in ref.~\cite{Rosso:2018acz}, for static black holes, right--moving paths only have positive heat flows. So the total heat flow $Q_H$ here only has a contribution from the isothermal path from $A$ to $B$, and $Q_C$ only from $C$ to $A$. Using $dQ=TdS$ and eq.~(\ref{eq:work-heat}) we find:
\begin{equation}\label{eq:engine-one}
\eta_{\rm upper}=1-
  \left(\frac{d-1}{d-2}\right)\frac{\Delta\left(TS\right)}{T_A\Delta S}\ ,
\end{equation}
where $\Delta X=X_C-X_A$. In a completely analogous way, we can design a partner engine that starts with a path going directly from $A$ to $C$ along the massless curve, then an isothermal compression from $C$ to $D$, and finally an adiabat/isochore from $D$ to $A$ (see fig.~\ref{fig:engine-one-lower}). Calculating the heat flows exactly the same way as for the previous engine, its efficiency is given by:
\begin{equation}\label{eq:engine-two}
\eta_{\rm lower}=
  1-\left(\frac{d-2}{d-1}\right)\frac{T_C\Delta S}{\Delta(TS)}\ .
\end{equation}

Since the highest and lowest temperatures these engine are in contact with are given by $T_A$ and $T_C$ respectively, their efficiency will be bounded by the efficiency of a Carnot engine working between those temperatures, $\eta_{\rm carnot}=1-T_C/T_A$. This is of course a statement of the Second Law of thermodynamics\footnote{The Second Law is being used differently here from how it was used in the  section~\ref{sec:RGFlow}. We are able to go from $A\to C$ in the second engine by using a high temperature reservoir to get inward heat flow. In the first engine, going from $C\to A$ is achieved with a low temperature reservoir. It is that latter process that is equivalent to the irreversible coarse--graining of an RG flow. (One way to imagine a field theory realization of heat flowing in is to let it become, by an appropriate coupling,  the reservoir of another field theory undergoing RG flow\cite{Johnson:2014yja}.)}.
We will discuss these bounds and their meaning in the next section. In fact, the engines of figure \ref{fig:engine} are two ``halves" of our Carnot engine, which is made from two isotherms and two adiabats/isochors (given by the cycle ABCD).

The reason for constructing these specific heat engines is the fact that their efficiency, heat flows and total work\footnote{The total work generated by one cycle of the engine can be calculated from $W=\eta Q_H$.}  can be written as the difference between temperatures and entropies between~$A$ and $C$. Since these two points lie on the massless curve, we will have an understanding of their meaning in the CFT.

\subsection{CFT Description of Holographic Heat Engines}
\label{sec:heat-engines-2}

We can now discuss the field theory meaning of the heat engines constructed in the previous section. For the massless path which the engines are partly constructed from, we know from the discussion in section~{\ref{sec:RGFlow}} that it is equivalent to a field theory RG flow for the upper engine. For the isothermal and isochoric/adiabatic paths we have considered when creating the engines  we are changing both the field theory and the state (see section \ref{sec:FirstLaw}). 

We now turn towards the field theory meaning of the work and heat flows of these engines, where we will be able to make  stronger statements. For Carnot's engine, these can be written as:
\begin{equation}\label{eq:30}
Q_{C/H}=T_{C/H}\Delta S\ ,
  \qquad \qquad
  W=-\Delta T\Delta S \ ,
\end{equation}
where remember that $\Delta X=X_C-X_A$. This is when the discussion of section \ref{sec:entropy} becomes useful: since all the quantities are evaluated on the massless curve, we can  directly translate the black hole temperature in terms of the radius and regulate the divergent black hole entropy according to eq.~(\ref{eq:entanglement}) so that we recover the entanglement entropy  of the ground state reduced to a ball in each CFT$_{A/C}$.  This means that we have an understanding of the field theory meaning of the heat flows and work for this engine; they are related to the difference in entanglement entropy between the CFTs. This same discussion holds for the work and heat flows of the  engines previously considered, where we get analogous expressions to those in eq.~(\ref{eq:30}).

Turning to the  efficiencies, notice that all appearances of temperatures involve only the ratio $T_C/T_A$, which we can write in terms of the RG scaling parameter $b$ as:
\begin{equation}\label{eq:23}
b=\frac{T_C}{T_A}=\frac{L_A}{L_C}=
  \left(
  \frac{a_d^{*(A)}}{a_d^{*(C)}}\right)^{1/(d-1)}\leq1 \ .
\end{equation}
We can now translate all the engine efficiencies we obtained in the previous section. First, for the Carnot engine we get:
\begin{equation}\label{eq:Car}
\eta_{\rm carnot}=1-\frac{T_C}{T_A}=
  1-b\ .
  \end{equation}
This function is possibly the simplest imaginable  dependence an efficiency could have on the RG scaling parameter $b$, remembering that it must vanish when $b=1$. Notice that it has relevance to engines (field theory tours) defined anywhere on the $(p,V)$ plane: Their efficiency is bounded by that of a Carnot engine whose efficiency can be written in this form, since all isotherms intersect (or asymptote to) the massless curve and hence connect to  a unique CFT.

For the efficiencies of the upper and lower engines (\ref{eq:engine-one}) and (\ref{eq:engine-two}) we get temperature and entropy ratios, which in principle should be regulated by the introduction of a cutoff according to eq.~(\ref{eq:cutoff}). However, the ratios  give a finite result once the cutoffs are taken away at the same rate {\it via} $\epsilon,\epsilon^\prime\to0$. Using eq.~(\ref{eq:temp-ent}) in (\ref{eq:engine-one}) and (\ref{eq:engine-two}) and taking the limit we find:
\begin{equation}\label{eq:11}
\eta_{\rm upper}=
  1-b\left[
  \left(\frac{d-1}{d-2}\right)\left(\frac{1-b^{d-2}}{1-b^{d-1}}\right)  
  \right]\ ,
  \quad {\rm and}\quad
  \eta_{\rm lower}=
  1-b\left[\left(\frac{d-2}{d-1}\right)\left(\frac{1-b^{1-d}}{1-b^{2-d}}\right)\right]\ .
\end{equation}
It is straightforward to show that the factors between square brackets are greater than or equal to unity in both cases, meaning that the efficiencies are bounded by Carnot's (\ref{eq:Car}). We then have that the efficiency is a cutoff independent quantity (despite the fact that the work and heat flows are not) which depends on the ratio of the number of degrees of freedom of the CFTs at $C$ and $A$ through $b$, the   RG scaling parameter\footnote{Another simple engine whose efficiency depends only on $b$ in the continuum limit is of the prototype rectangular form~\cite{Johnson:2014yja} using two isobars and two adiabats, with CFTs $A$ and $C$  on diagonally opposite corners. The efficiency (evaluated using just  mass differences~\cite{Johnson:2016pfa}) turns out to be $\eta=1-b^d$,   bounded above by the Carnot efficiency which in this case is the more complicated expression $\eta_{\rm carnot}=1-b[b^2d-(d-2)]/[d-(d-2)b^2]$. We can write the latter as $1-{\tilde b}$, where ${\tilde b}\leq b^d$ is the RG scaling parameter relating CFTs at $A^\prime$ and $C^\prime$ located further up and down the $M=0$ curve. (Note that in this case  $1\geq b\geq[(d-2)/d]^{1/2}$, to avoid having $T<0$ on the cycle.)}. 

A few special limits are of interest. The limit of small engine cycles  ($\xi = 1-b\ll 1$) is appealing, since it involves paths that do not deviate too far from the massless curve. Expanding eq.~(\ref{eq:11}) around small $\xi$ we find $\eta_{\rm upper} =\xi/2+(d/12)\xi^2+\cdots$ and $\eta_{\rm lower} =\xi/2+((3-d)/12)\xi^2+\cdots$, where notice that $\eta_{\rm carnot}=\xi$. We then find that to leading order, both engines are half as efficient as Carnot\footnote{Notice that in the interesting case of $d=3$  the efficiencies are exactly $\eta_{\rm lower}=\xi/2$ and $\eta_{\rm upper}=\xi/(2-\xi)$.}. On the other hand, if $b$ is small, corresponding to a large cycle, to leading order we get  $\eta_{\rm upper} \rightarrow 1- bf(d)$ and $\eta_{\rm lower} \rightarrow 1- f(d)^{-1}$.  Finally, it is interesting to consider the large~$d$ limit, where to leading order we get $\eta_{\rm upper} \rightarrow  \eta_{\rm carnot}$ and $\eta_{\rm lower} \rightarrow 0$.

\section{Conclusions}
\label{sec:Conclusion}

In this work we have explored the rich extended thermodynamics of hyperbolic black holes in anti--de Sitter spacetime, and revealed its fascinating connection with the properties of conformal field theories. This is a natural extension of the observations of ref. \cite{Casini:2011kv}, where the core object is a  hyperbolic slicing of AdS that has an interpretation as a special black hole. By considering that special black hole as just part of a larger  extended thermodynamics, where other black holes join the physics, we enlarged the framework and have found a long sought--after setting in which the extended thermodynamics of black holes has a clear application to matters in holographic descriptions of quantum field theories. Moreover, holographic heat engines and their efficiency have now been cast more clearly in field theory terms, along the lines of the conjectures made in ref.~\cite{Johnson:2014yja}. 

In making the connection between these two areas, certain structures that were merely analogous become precisely equivalent. Most intriguingly, moving along the curve of  massless  black holes in the $(p,V)$ plane towards higher pressure is the description of an RG flow from the UV to IR, and the irreversibility of coarse--graining is a consequence of the Second Law of thermodynamics. This joins the understanding of the First Law (already partly uncovered in the literature\cite{Blanco:2013joa,Faulkner:2013ica,Kastor:2014dra}) as expressing the changes to the CFT due to perturbations (entering through changes to entanglement) and flows (changing the number of degrees of freedom). From that, given that there is a perturbation from the massless hyperbolic holes to negative mass, we were able to capture the relation between negative energy density and entanglement, which was previously studied in the literature but from a very different perspective \cite{Blanco:2013lea,Bianchi:2014qua,Blanco:2017akw}. It would be interesting to further explore this connection from this angle.

For the three interesting holographic heat engines we studied, we could express  their heat flows and work entirely in field theory terms as the difference in the entanglement entropy  of two CFTs visited along the cycle. Their efficiencies, which have a natural bound (another consequence of the Second Law of thermodynamics) supplied by the Carnot efficiency, are simply functions of the ratio of degrees of freedom of those two CFTs. The Carnot efficiency is, in a sense, the simplest function of all, and seems to naturally express the limit on any other engine (no matter where it is defined in the $(p,V)$ plane) in terms of two CFTs, which is intriguing. 

Now that they can be expressed in terms of field theory quantities, it would be interesting to explore further what  holographic heat engines can teach us about the CFTs and their deformations, and what the bounds on efficiencies  might mean. This could also lead to applications of the broader work that has been recently done to understand how to compute general black hole engine efficiencies, and make meaningful comparisons among them\cite{Johnson:2016pfa,Chakraborty:2016ssb,Hennigar:2017apu,Chakraborty:2017weq,Rosso:2018acz}.

A possible clue to more applications is the fact that the efficiencies that we computed have rather  universal properties. In ref.~\cite{Rosso:2018acz} it was shown that (at least for static black holes), there is a large set of equivalent engines that one can explicitly construct that will result in the same efficiency formulae. How this equivalency plays out in terms of the CFTs is worth exploring in future work.

Finally, we noted how known  holographic RG flows between CFTs  (which have their own detailed gravity description) are accounted for in moves up the massless curve in the $(p,V)$ plane. A natural conjecture is that there are other pressure--increasing moves   corresponding  to  other holographic RG flows known in the literature. An exploration of this possibility (for example by matching asymptotic geometries) could be very useful for precisely mapping out the CFT physics of the $(p,V)$ plane.

\section*{Acknowledgements}
The work of CVJ and FR was funded by the US Department of Energy  under grant DE-SC 0011687. FR would like to thank Hubert Saleur and Namit Anand for useful discussions at the beginning of this work. CVJ would  like to thank the Aspen Center for Physics for hospitality during the early stages of this project, and Amelia for her support and patience.

\bibliographystyle{utphys}
\bibliography{heat_engine_entanglement}

\end{document}